\documentclass[12pt]{article}
\def\dddot#1{\mathinner{\buildrel\vbox{\kern5pt\hbox{...}}\over{#1}}}
\def\ddddot#1{\mathinner{\buildrel\vbox{\kern5pt\hbox{....}}\over{#1}}}

\def\d{\mbox{\rm d}}
\usepackage{exscale}\usepackage{graphicx}

\begin{document}
\title{$\lambda$-symmetries and Jacobi Last Multiplier}

\author{M.C. Nucci\footnote{e-mail: nucci@unipg.it}\\
 Dipartimento di Matematica
e Informatica,\\ Universit\`a degli Studi di Perugia \& INFN Sezione Perugia,\\
06123 Perugia,
Italy\\[0.5cm]
D. Levi\footnote{email: levi@roma3.infn.it}\\
Dipartimento di Ingegneria Elettronica,\\ Universit\`a degli Studi Roma Tre \&
INFN Sezione Roma TRE,\\ Via della Vasca Navale 84, 00146 Roma, Italy}

 \maketitle
\begin{abstract}
We show that  $\lambda$-symmetries can be algorithmically obtained by using the
Jacobi last multiplier. Several examples are provided.

\end{abstract}
\section{Introduction}
 Lie group analysis is  the most powerful general tool to find the solution
of Ordinary Differential Equations (ODEs). However an ODE of $n$th order does
not always admit Lie point symmetries. Moreover this Lie group analysis  is
useless when applied to $n$ equations of first order because they admit an
infinite number of symmetries, and there is no systematic way to find even a
one-dimensional Lie symmetry algebra. One may try to derive an admitted
$n$-dimensional solvable Lie symmetry algebra by making an ansatz on the form
of its generators but when successful (rarely) it is just a lucky guess.
However, in \cite{kepler} it has been shown that any system of $n$ equations of
first order can be transformed into an equivalent system where at least one of
the equations is of second order. Then the admitted Lie symmetry algebra is no
longer infinite-dimensional, and nontrivial symmetries of the original system
may be retrieved. This idea has been successfully applied in several instances
(\cite{kepler}, \cite{valenu}, \cite{harmony}, \cite{lorpoin}, \cite{MICZ},
\cite{goldfish}, \cite{core}, \cite{gradnuc1}).\\
In \cite{jlm05}  another method was devised. It uses the Jacobi last multiplier
(\cite{Jacobi 44 a}, \cite{Jacobi 44 b}, \cite{Jacobi 45 a}, \cite{JacobiVD},
\cite{Lie1874}, \cite{Lie 12 a}, \cite{Bianchi 18 a}, \cite{Whittaker}) to
transform a system of $n$ first-order equations into an equivalent
 system of $n$ equations where  one of the equations is of
second order, namely the order of the system is raised by one. In \cite{jlm05},
among other examples, the method was successfully applied to the second-order
equation \cite {Kamke 83}[Ch. 6, 542{\it ff}]:
\begin{equation}y''={y'^2\over y} +f'(t)
y^{p+1}+pf(t)y'y^p,\label{kamkeq}
\end{equation}
where $p\neq 0$ is a real constant and $f\neq 0$ is an arbitrary function of
the independent variable $t$. This equation does not possess Lie point
symmetries for general $f(t)$ and yet is trivially integrable
\cite{gonzalez88}. In \cite{jlm05} the introduction of the Jacobi last
multiplier led to an equivalent system of two equations, one of first order and
the other of second order, which admits enough Lie symmetries in
order to integrate it by quadrature. \\
In  \cite{MurRo01} and \cite{MurRo03}  Muriel and Romero introduced the
so-called $\lambda$-symmetries which were later included into the telescopic
symmetries by  Pucci and Saccomandi \cite{PuSa02}. Again as in the case of Lie
point symmetries of first-order equations the real problem is to find solutions
of the determining equations. This problem has been clearly stated by Pucci and
Saccomandi  on page 6154 of their article. \\
In  \cite{CGM04} and \cite{GM04} Cicogna, Gaeta and Morando provided a
geometrical characterization of $\lambda$-symmetries for both ordinary and
partial differential equations. However, in either paper they did not address
the problem of solving the determining equation,
 they just obtain a $\lambda$-symmetry by guesswork.\\
In  \cite{Ferraioli} Catalano Ferraioli showed that $\lambda$-symmetries
correspond to a special type of nonlocal symmetries and in a final remark
observed that also telescopic symmetries can be recovered by nonlocal
symmetries. Yet there is much guesswork involved in order to find nonlocal
symmetries, and on page 5485, Remark 2, Catalano Ferraioli  wrote: ``However,
we note that the problem of determining the general solution of
(11)\footnote{It is a linear first-order partial differential equation.},
should be at least as difficult as solving the given ODE. Therefore in practice
it could not be so easy to determine such correspondence
(see example 3)". \\
In \cite{PInceeq} the method  described in \cite{jlm05} was applied to example
3 in \cite{Ferraioli}, i.e
\begin{equation}
y''=\frac{y'^2}{y}+\left(y+\frac{t}{y}\right)y'-1,\label{PXIVpeq}
\end{equation}
that is a second-order Painlev\'e-type equation \cite{Painl1900},  which does
not admit any Lie point symmetries and is a particular case\footnote{It
corresponds to assume $Q(t)=1$ and $S(t)=t$.} of Painlev\'e XIV equation
\cite{Ince}, i.e.:
\begin{equation}
y''=\frac{y'^2}{y}+\left(Q(t)y+\frac{S(t)}{y}\right)y'+Q'(t)y^2-S'(t),
\label{PXIVeq}
\end{equation}
which does not possess any Lie point symmetry for arbitrary $Q(t), S(t)$
although it has a Riccati-type first integral, i.e.:
\begin{equation}
\frac{y'-Q(t)y^2+S(t)}{y}=a_1, \label{fintPXIV}
\end{equation}
with $a_1$ an arbitrary constant. In \cite{PInceeq} the introduction of the
Jacobi last multiplier into equation (\ref{PXIVpeq}) led to a system of three
equations, two of first order and one of second order. This system admits a
three-dimensional solvable Lie symmetry algebra and therefore can be reduced to
a Riccati equation that can be integrated in terms of Airy functions. Thus  a
new first integral different from (\ref{fintPXIV}) was found, i.e.
\begin{equation}
a_2=\frac{(y^2 + y' + t){\rm AiryAi}\left(\xi\right)+2y{\rm
AiryAi}\left(1,\xi\right)}{(y^2 + y' + t){\rm AiryBi}\left(\xi\right)+2y{\rm
AiryBi}\left(1,\xi\right)}, \quad \quad
 \xi={\displaystyle
\frac{y^4+2(t-y')y^2+(t+y')^2}{4y^2}}. \label{fint2PXIVp}
\end{equation}
with  $a_2$ an arbitrary constant. Then, combining the two first integrals
(\ref{fintPXIV})-- with $Q(t)=1$ and $S(t)=t$ -- and (\ref{fint2PXIVp}), one
gets
 the general solution of equation (\ref{PXIVpeq}) in implicit form,
i.e.
\begin{equation}
\frac{(2y+a_1){\rm AiryAi}\left(\xi\right)+2{\rm
AiryAi}\left(1,\xi\right)}{(2y+a_1){\rm AiryBi}\left(\xi\right)+2{\rm
AiryBi}\left(1,\xi\right)}=a_2, \quad \quad
 \xi=t+{\displaystyle \frac{a_1^2}{4}}.
\end{equation}
 This suggests that there might be a direct link between the
Jacobi last multiplier and $\lambda$-symmetries.

In the present paper we show that the introduction of the Jacobi last
multiplier allows to find $\lambda$-symmetries algorithmically.

 In the next Section, we briefly recall the connection between nonlocal
symmetries and $\lambda$-symmetries as shown by Catalano Ferraioli in
\cite{Ferraioli}, and some essential properties of the Jacobi last
 multiplier. Then we show how to use the Jacobi last multiplier
 to find $\lambda$-symmetries. Recently in \cite{MurRo_W10} Muriel and Romero
 have classified all the $\lambda$-symmetries of any second-order ODE through
 an equivalence relationship and proved that two $\lambda$-symmetries lead to
 functionally independent first integral if and only if they are in different
 equivalence classes. This important result allows us to discriminate
  the more than one $\lambda$-symmetry that we found in some of the examples.
 In Section 3 we apply our method based on the Jacobi last multiplier and
 either recover known or find new $\lambda$-symmetries of several
  differential equations of second order:
 equation (\ref{kamkeq}) as given in \cite{gonzalez88}, equations Painlev\'e V,
XIV (\ref{PXIVeq}), XV, and XVI as given in \cite{Ince}\footnote{In
 particular we recover the known first integrals \cite{Ince}.}, and examples 4 and 5
as given in \cite{CatalanoMorando09}. Section 4 is devoted to some conclusions.

\section{$\lambda$-symmetries and Jacobi Last Multiplier}

Let us consider an $n^{th}$-order ODE:
\begin{equation}
 y^{(n)}=f(t,y,y',y'',\ldots,y^{(n-1)}),\label{node} \end{equation}
 where by an apex we mean the order of differentiation.
 In \cite{Ferraioli}  a nonlocal interpretation
 of $\lambda$-symmetries was given.
 There Catalano Ferraioli has shown that seeking $\lambda$-symmetries of eq.
 (\ref{node}) is equivalent  to add
to  equation (\ref{node}) the equation
\begin{equation}\omega'=\lambda
\label{weq}\end{equation} for the new field $\omega(t)$ with
$\lambda=\lambda(t,y,y',y'',\ldots,y^{(n-1)})$. The symmetries of (\ref{node},
\ref{weq}) are obtained by considering the infinitesimal  operator:
\begin{equation}
Y=\tilde \tau(t,y,\omega)\partial_t+\tilde \eta(t,y,\omega)\partial_y+\tilde
\xi(t,y,y',y'',\ldots,y^{(n-1)}, \omega)\partial_{\omega} \label{ig1}
\end{equation}
with the constraint
\begin{equation} \label{ig2}
[Y,\partial_{\omega}]=Y.
\end{equation}
Under the constraint (\ref{ig2}) we get $Y=e^{\omega} [X+ \xi
\partial_{\omega}]$ where
\begin{equation} \label{ig3}
X=\tau(t,y)\partial_t+ \eta(t,y)\partial_y
\end{equation}
and $\xi=\xi(t,y,y',y'',\ldots,y^{(n-1)})$. The prolongation of $X$ is
\begin{eqnarray}
\mbox{pr}X&=&\tau(t,y)\partial_t+ \eta(t,y)\partial_y
+\eta^{(1)}(t,y,y',y'',\ldots,y^{(n-1)})\partial_{y'}\nonumber\\&&+\eta^{(2)}(t,y,y',y'',\ldots,y^{(n-1)})\partial_{y''}
+ \cdots \label{Xpr1}
\end{eqnarray}
with
\begin{equation} \label{Xpr2}
\eta^{(n+1)}= \Big[(D_t+\lambda)\eta^{(n)}-y'(D_t+\lambda)\tau\Big]
\end{equation}
where $D_t=\partial_t+\sum_{k=0}^{n}y^{(k+1)}\partial_{y^{(k)}}$,
$y^{(0)}\equiv y$, and $\eta^{(0)}\equiv \eta$. A $\lambda$-symmetry of
equation
 (\ref{node}) is any solution of the determining equation
\begin{equation}
\mbox{pr}X\left(y^{(n)}-f(t,y,y',y'',\ldots,y^{(n-1)}\right)
{|}_{_{y^{(n)}=f}}=0, \label{Xndeteq}
\end{equation}
depending on the three unknowns  $\lambda$, $\tau$ and $\eta$ and therefore is
 highly
 undetermined. In \cite{MurRo_W10} proved that any $\lambda$-symmetry can be put in an
evolution form as a translation in $y$, i.e. with $\tau=0$ and $\eta=1$. Hence
equation (\ref{Xndeteq}) becomes determined  since it has only one unknown,
 but it is still difficult to solve it since it is a nonlinear
partial differential equation in the unknown $\lambda$.
\\\\ If equation
(\ref{node}) is transformed into an equivalent system of first-order equations,
i.e.
\begin{equation}{w}_i' = W_i(t,w_1,\dots,w_n), \label{wisis} \end{equation}
then its Jacobi last multiplier $M$ is obtained by solving the following
differential equation
\begin {equation}
 \frac {\d\log (M)} {\d t} + \sum_{i=1}^n\frac {\partial W_i}
{\partial w_i}=0, \label {Meq}
\end {equation}
i.e.
\begin{equation}
M=\exp\left(-\int \sum_{i=1}^n\frac {\partial W_i} {\partial
w_i}\,dt\right).\label{Mint}\end{equation} In \cite{jlm05}  all the properties
of the Jacobi last multiplier are listed.   There three strategies were
proposed  with the purpose of finding Lie symmetries of any system
(\ref{wisis}):
\begin{enumerate}

\item Eliminate -- if possible -- each of the variables $w_i$  in order to obtain
an equivalent $n$-order system  which contains  a single equation of second
order and $n-2$ equations of first order. The admitted Lie symmetry algebra is
no longer infinite-dimensional and the Lie group analysis can be usefully
applied (\cite{kepler}, \cite{valenu}, \cite{core}). From this strategy we also
get first integrals (\cite{marcelnuc}, \cite{lorpoin}, \cite{MICZ}).

\item Decrease the order of system (\ref{wisis}) by one choosing
one of the variables $w_i$ as the new independent variable. Then apply either
Strategy 1 or 3 in this list. For example, if $w_1\equiv y$ is the new
independent variable, then system (\ref{wisis}) becomes
\begin{equation}
\frac{{\rm d} w_k}{{\rm d} y}=\frac{W_{k}}{W_1}
\equiv\Omega_k(y,w_2,\ldots,w_n)\quad\quad (k=2,n) \label{wisisr}
\end{equation}
This method has been applied in several examples (e.g. \cite{harmony},
\cite{MICZ}, \cite{goldfish}) since its first instance, the Kepler problem
\cite{kepler}.

\item Increase the order  using the transformation suggested by the Jacobi last
multiplier, i.e. introducing a new dependent variable $R$ such that
\begin{equation}
\frac{{\rm d} R}{{\rm d} t}= \sum_{i=1}^n\frac {\partial W_i} {\partial
w_i}\label{Req}
\end{equation}
and eliminating  -- if possible -- each $w_i$ which appears in (\ref{Req}).
Then system (\ref{wisis}) reduces to a system given by a single second-order
ODE and $n-1$ first-order equations. If a new independent variable was chosen
(Strategy 2), say $w_1\equiv y$, then $R$ will satisfy the following equation:
\begin{equation} \label{Req1}
\frac{{\rm d} R}{{\rm d} y}=\sum_{k=2}^n\frac {\partial \Omega_k} {\partial
w_k}
\end{equation}
and then the strategy goes as above.
\end{enumerate}

In \cite{jlm05} Strategy 3 was applied to find Lie symmetries of several
systems.\\\\
The similarity between Strategy 3 and the nonlocal approach to
 $\lambda$-symmetries as given by Catalano Ferraioli suggests to search  for
$\lambda$-symmetries such that
\begin{equation}
\omega'=\lambda=\sum_{i=1}^n\frac {\partial W_i} {\partial w_i}.
\label{lambdaJ}
\end{equation}
This implies, when feasible, that $\omega=\log(1/M)$. In fact this connection
cannot be made if the divergence of the system (\ref{wisis}), namely
$Div\equiv\sum_{i=1}^n\frac {\partial W_i} {\partial w_i}$, is zero, since then
any Jacobi last multiplier is a first integral of (\ref{wisis}) and therefore
such is $\omega$. \\However there are many ways in which a system of $n$
first-order equations can be written as a single equation of $n$th order and
viceversa. The Jacobi last multiplier is then different and if one way yields
$Div=0$, another way may yield $Div\neq 0$. In particular the following system
of two first order equations
\begin{equation}
w_1'=W_1(t,w_2),\quad\quad\quad w_2'=W_2(t,w_1)
\end{equation}
has $Div=0$ and therefore $M_{[w_1,w_2]}=1$ is one of its Jacobi last
multipliers. If we derive $w_1$ from the second equation, i.e. $w_1=\overline
W_2(t,w_2')$, then an equivalent second-order ODE is obtained, i.e.
\begin{equation}
w_2''=\frac{W_1(t,w_2)-\displaystyle\frac{\partial }{\partial t }\overline
W_2(t,w_2')}{\displaystyle\frac{\partial }{\partial {w_2'}}\overline
W_2(t,w_2')}
\end{equation}
which has $Div\neq 0$ since\footnote{This is one of the known properties of the
Jacobi last multiplier \cite{Bianchi 18 a}, \cite{jlm05}: given a non-singular
transformation of variables
$$
\tau:\quad(w_1,w_2,\ldots,w_n)\longrightarrow(r_1,r_2,\ldots,r_n),
$$
\noindent then the last multiplier $M_{[r]}$ of the new system $
r_i'=R_i(t,r_1,\ldots,r_n)$ is given by:
$$
M_{[r]}=M_{[w]}\frac{\partial(w_1,w_2,\ldots,w_n)}{\partial(r_1,r_2,\ldots,r_n)}\,.
$$
}
\begin{equation}
M_{[w_2]}=M_{[w_1,w_2]}\frac{\partial(w_1,w_2)}{\partial(w_2,w_2')}
=M_{[w_1,w_2]}\left
|
\begin {array} {cc}
0& \displaystyle\frac{\partial }{\partial {w_2'}}\overline W_2(t,w_2')\\
[0.3cm] 1 & 0
\end {array}\right|= \displaystyle\frac{\partial }{\partial {w_2'}}\overline
W_2(t,w_2'),
\end{equation}
i.e.  its Jacobi last multiplier cannot be a constant. An illustrative example
of such an instance is the following system studied in \cite{nuctam_2lag}
\begin{equation} r_1'= b\exp(r_2) + a \quad\quad\quad
 r_2' = B\exp(r_1) + A\label{VLt}
\end{equation}
which has obviously $Div=0$. Following \cite{nuctam_2lag} we can transform
system (\ref{VLt}) into an equivalent second-order ordinary differential
equation by eliminating $r_1$. In fact from the second equation in (\ref{VLt})
one gets
\begin {equation}r_1=\log\left(\frac{r_2' - A}{B}\right),\label{r1}\end{equation}
and the equivalent second-order equation in $r_2$ is the following
\begin {equation}
 r_2''=- \Big(b\exp(r_2)+a\Big)(A- r_2'). \label{VLr2}
\end {equation}
which has $Div= b\exp(r_2) + a\neq 0$.
\\\\
As a final remark we recall that recently \cite{MurRo_W10} Muriel and Romero
have proved the equivalence between two $\lambda$-symmetries  of a second-order
ordinary differential equation
\begin{equation}y''=\phi(t,y,y').\label{2ode}\end{equation} Assuming that
$\lambda_1$ and $\lambda_2$ yield two $\lambda$-symmetries
$X_1=\tau_1(t,y)\partial_t+\eta_1(t,y)\partial_y$ and
$X_2=\tau_2(t,y)\partial_t+\eta_2(t,y)\partial_y$ of equation (\ref{2ode}),
respectively, they are equivalent if and only if \begin{equation}
 Q_1(A+\lambda_2)(Q_2)-Q_2(A+\lambda_1)(Q_1)=0,       \label{lamequiv}
\end{equation}
where $Q_i=\eta_i-y'\tau_i, \;(i=1,2)$ and
$A=\partial_t+y'\partial_y+\phi(t,y,y')\partial_{y'}$ is the vector field
associated with (\ref{2ode}). This equivalence will be used in the next Section
3 when we obtain more than one $\lambda$-symmetry.
\section{Examples}
In the following examples we denote by $\lambda_{k}$ the $\lambda$-symmetries
presented by Muriel, Romero, Catalano Ferraioli and Morando in the references
\cite{MurRo01}, \cite{MurRo03}, \cite{CatalanoMorando09} and by $\lambda_J$ the
$\lambda$-symmetry that we find by using formula (\ref{lambdaJ}) once we
rewrite (\ref{node}) as (\ref{wisis}).

\subsection{Equation (\ref{kamkeq})}
In \cite{MurRo01} a $\lambda$-symmetry of equation (\ref{kamkeq}) was
determined, i.e.\begin{equation} X^{(\lambda_k)}= \partial_y \quad {\rm
with}\quad \lambda_{k}= py^pf(t)+y'/y. \label{kamlamk}\end{equation} The
divergence of equation (\ref{kamkeq}) yields
\begin{equation}
\lambda_{J}=py^pf(t)+2\frac{y'}{y}. \label{kamlamJ}
\end{equation}
 If we put $\lambda_{J}$ into
(\ref{Xpr1}) then the solution of the determining equations (\ref{Xndeteq})
yields two $\lambda$-symmetries, i.e.
\begin{equation}
X^{(\lambda)}_1=\frac{1}{y}\partial_y,\quad\quad
X^{(\lambda)}_2=\frac{1}{y^2}\partial_t+y^{p-1}f(t)\partial_y. \label{kamsimj}
\end{equation}
The first prolongation of $X^{(\lambda)}_1$, i.e.
\begin{equation}
\mbox{pr}X^{(\lambda)}_1=X^{(\lambda)}_1+\left(py^{p-1}f(t)+\frac{y'}{y^2}\right)\partial_{y'}
\end{equation}
yields the first-order invariants
\begin{equation} y_1=-y^pf(t)+\frac{y'}{y},\quad \quad
t_1=t\end{equation} that replaced into equation (\ref{kamkeq}) generate the
first-order equation
\begin{equation}
 \frac{d y_1}{d t_1}=y'=0,
\end{equation}
as obtained in \cite{MurRo01}. The first prolongation of $X^{(\lambda)}_2$,
i.e.
\begin{equation}
\mbox{pr}X^{(\lambda)}_2=X^{(\lambda)}_2+\left(y^{2p-1}f^2(t)p
+y^{p-1}f'(t)+y^{p-2}y' f(t)\right)\partial_{y'}
\end{equation}
yields the first-order invariants
\begin{equation} y_2=F^{\frac{1}{p}}(t)y' + \frac{F'(t)}{pF(t)},\quad \quad t_2=
\frac{1}{y^p}+p\int f(t)\,{\rm d}t\end{equation} where $F(t)=t_2 - p\int
f(t)\,{\rm d}t$, that replaced into equation (\ref{kamkeq})
generate the same invariant as $X^{(\lambda)}_1$.\\
 This result should not be a
surprise since the two $\lambda$-symmetries (\ref{kamsimj}) and the
$\lambda$-symmetry $\lambda_k$ found in \cite{MurRo01} are all equivalent  as
defined by Muriel and Romero in \cite{MurRo_W10}. In fact substituting
$X^{(\lambda)}_1$ and $X^{(\lambda)}_2$ -- that have  the same $\lambda_J$
(\ref{kamlamJ}) -- into (\ref{lamequiv}) yields:
\begin{eqnarray}
Q_1(A+\lambda_J)(Q_2)-Q_2(A+\lambda_J)(Q_1)=Q_1(A)(Q_2)-Q_2(A)(Q_1)&\nonumber\\
=\frac{1}{y}\left[ y^{p-1}f'(t)+y'\left(
(p-1)y^{p-2}f(t)+2\frac{y'}{y^3}\right)-\frac{1}{y^2} \left({y'^2\over y}
+f'(t) y^{p+1}+p f(t)y'y^p\right) \right ]&\nonumber\\+\left(y^{p-1}f(t)-
\frac{y'}{y^2}\right)\frac{y'}{y^2}=0&
\end{eqnarray}
since $Q_1=1/y$  and $Q_2=y^{p-1}f(t)-y'/y^2$. Also substituting
$X^{(\lambda)}_1$  and $X^{(\lambda_k)}$  and their corresponding $\lambda_J$
and $\lambda_k$ into (\ref{lamequiv}) yields:
\begin{equation}
Q_1(A+\lambda_k)(Q_k)-Q_k(A+\lambda_J)(Q_1)=-\frac{1}{y}\left(
py^pf(t)+\frac{y'}{y}\right)+\frac{1}{y} \left(py^pf(t)+2\frac{y'}{y} \right)
-\frac{y'}{y^2}=0
\end{equation}
since $Q_k=1$.

\subsection{Painlev\'e-Ince V}
The divergence of  Painlev\'e-Ince equation V
\begin{equation}
y''=-2yy'+q(t) y'+q'(t)y \label{PVeq}
\end{equation} yields
\begin{equation} \lambda_J=-2y+q(t).\end{equation}
If we put $\lambda_{J}$ into the $\lambda$--prolongation then (\ref{Xndeteq})
yields two $\lambda$-symmetries, i.e.
\begin{equation}
X^{(\lambda)}_1=\partial_y,\quad\quad
X^{(\lambda)}_2=\partial_t+\left(yq(t)-y^2\right)\partial_y. \label{PVlamsym}
\end{equation}
These two $\lambda$-symmetries (\ref{PVlamsym}) are equivalent
\cite{MurRo_W10}. In fact
\begin{equation}
Q_1(A)(Q_2)-Q_2(A)(Q_1)=yq'(t)+y'\left(q(t)-2y\right)-\left(-2yy'+q(t)y'+q'(t)y\right)=0
\end{equation}
since $Q_1=1$  and $Q_2=yq(t)-y^2-y'$.\\
Therefore we consider only one $\lambda$-symmetry, $X^{(\lambda)}_1$. Its first
prolongation,  i.e.
\begin{equation}
\mbox{pr}X^{(\lambda)}_1=X^{(\lambda)}_1+\left(q(t)-2y\right)\partial_{y'}
\end{equation}
yields the first-order invariants
\begin{equation}  y_1=-yq(t)+y^2+y',\quad \quad
t_1=t\end{equation} that replaced into equation (\ref{PVeq}) generate the
first-order equation
\begin{equation}
y_1 '=0 \Longrightarrow y_1=a_1 \Longrightarrow -yq(t)+y^2+y'=a_1,
\label{fintPV}
\end{equation}
and therefore the known first integral of (\ref{PVeq}) is derived
\cite{Ince}.\\
As far as we know $X^{(\lambda)}_1$ in (\ref{PVlamsym}) is a new
$\lambda$-symmetry of (\ref{PVeq}).

\subsection{Painlev\'e-Ince XIV}
In \cite{MurRo09} a $\lambda$-symmetry of equation Painlev\'e-Ince  XIV
equation (\ref{PXIVeq}) was determined by assuming that $\lambda$ was linear
with respect to $y'$, and thus the following $\lambda$-symmetry was found
\begin{equation} X^{(\lambda_k)}=
\partial_y \quad {\rm with}\quad \lambda_{k}=
yQ(t)+\frac{S(t)}{y}+\frac{y'}{y}. \label{PXIVlamk}\end{equation}
 Instead the divergence of  equation (\ref{PXIVeq}) yields
\begin{equation}
 \lambda_J=\frac{S(t)}{y}+Q(t)y+D_t\left(\log\left(y^2\right)\right).\end{equation}
If we put $\lambda_{J}$ into the $\lambda$--prolongation and solve
 the determining equation  (\ref{Xndeteq}) we get two
$\lambda$-symmetries, i.e.
\begin{equation}
X^{(\lambda)}_1=\frac{1}{y}\partial_y,\quad\quad
X^{(\lambda)}_2=\frac{1}{y^2}\partial_t+\left(-\frac{S(t)}{y^2}+Q(t)\right)\partial_y.
\end{equation}
One can prove that these two $\lambda$-symmetries and that found  in
\cite{MurRo09} are equivalent. \\The first prolongation of $X^{(\lambda)}_1$,
i.e.
\begin{equation}
\mbox{pr}X^{(\lambda)}_1=X^{(\lambda)}_1+\left(\frac{S(t)}{y^2}+Q(t)+\frac{y'}{y^2}\right)\partial_{y'}
\end{equation}
yields the first-order invariants
\begin{equation}  y_1=\frac{S(t)}{y}-Q(t)y+\frac{y'}{y},\quad \quad
t_1=t\end{equation} that replaced into equation (\ref{PXIVeq}) generate the
first-order equation
\begin{equation}
y_1'=0 \Longrightarrow y_1=a_1 \Longrightarrow
\frac{S(t)}{y}-Q(t)y+\frac{y'}{y}=a_1,
\end{equation}
and thus the known first integral (\ref{fintPXIV}) of (\ref{PXIVeq}) is
derived.

\subsection{Painlev\'e-Ince XV}
 Painlev\'e-Ince XV equation
  \begin{equation}
  y''=\frac{y'^2}{y}+\frac{y'}{y}+r(t)y^2-y\frac{d}{dt}\left(
  \frac{r'(t)}{r(t)}\right) \label{PXVeq}\end{equation}
is known to possess a first integral \cite{Ince}, i.e.:
\begin{equation}\frac{1}{y^2}\left(\frac{r'(t)}{r(t)}y+y'+1\right)^2-2\left(r(t)y
+\int{r(t) dt}\right)=a_1. \label{fintPXV}\end{equation}
   The divergence of Painlev\'e-Ince  XV equation (\ref{PXVeq}) yields
\begin{equation}
 \lambda_J=\frac{1}{y}+D_t\left(\log\left(y^2\right)\right).\end{equation}
If we put $\lambda_{J}$ into (\ref{Xpr1}) then (\ref{Xndeteq}) yields one
$\lambda$-symmetry, i.e.
\begin{equation}
X^{(\lambda)}=\frac{1}{y^2}\,\partial_t-\frac{r'(t)y+r(t)}{r(t)y^2}\,\partial_y.
\end{equation}
The first prolongation of $X^{(\lambda)}$, i.e.
\begin{equation}
\mbox{pr}X^{(\lambda)}=X^{(\lambda)}+\frac{-r''(t)r(t)y^2+r'(t)^2y^2-r'(t)r(t)y(y'+1)-
r(t)^2(y'+1)}{r(t)^2y^3}\,\partial_{y'}
\end{equation}
yields the first-order invariants
\begin{equation}  \tilde y=\frac{r'(t)}{r(t)}+\frac{y'+1}{y},\quad \quad
\tilde t=r(t)y+\int r(t) dt\end{equation} that replaced into equation
(\ref{PXVeq}) generate the first-order equation
\begin{equation}
\frac{{\rm d}\tilde y}{{\rm d} \tilde t}=\frac{1}{\tilde y} \Longrightarrow
\tilde y^2-2\tilde t=a_1 \Longrightarrow
\left(\frac{r'(t)}{r(t)}+\frac{y'+1}{y}\right)^2-2\left(r(t)y+\int r(t)
dt\right)=a_1,
\end{equation}
and thus the known first integral (\ref{fintPXV}) of (\ref{PXVeq}) is
derived.\\
As far as we know $X^{(\lambda)}$ is a
novel $\lambda$-symmetry of equation (\ref{PXVIeq}).\\

\subsection{Painlev\'e-Ince XVI}
Painlev\'e-Ince XVI equation
  \begin{equation}
  y''=\frac{y'^2}{y}-q'(t)\frac{y'}{y}+y^3-q(t)y^2+q''(t) \label{PXVIeq}\end{equation}
is known to possess a first integral \cite{Ince}, i.e.:
\begin{equation}\left(\frac{y'-q'(t)}{y}\right)^2-(y-q(t))^2=a_1. \label{fintPXVI}\end{equation}
   The divergence of Painlev\'e-Ince  XVI equation (\ref{PXVIeq}) yields
\begin{equation}
 \lambda_J=-\frac{q'(t)}{y}+D_t\left(\log\left(y^2\right)\right).\end{equation}
 If we put $\lambda_{J}$ into (\ref{Xpr1}) then (\ref{Xndeteq}) yields one
$\lambda$-symmetry, i.e.
\begin{equation}
X^{(\lambda)}=\frac{1}{y^2}\,\partial_t+\frac{q'(t)}{y^2}\,\partial_y.
\end{equation}
 The first prolongation of $X^{(\lambda)}$,
(\ref{Xpr1}), i.e.
\begin{equation}
\mbox{pr}X^{(\lambda)}=X^{(\lambda)}+\frac{q''(t)y-q'(t)^2+q'(t)y'}{y^3}\,\partial_{y'}
\end{equation}
 yields the first-order
invariants
\begin{equation}  \tilde y=\frac{y'-q'(t)}{y},\quad \quad
\tilde t=y-q(t)\end{equation}
 that replaced into equation
(\ref{PXVIeq}) generate the first-order equation
\begin{equation}
\frac{{\rm d}\tilde y}{{\rm d} \tilde t}=\frac{\tilde t}{\tilde y}
\Longrightarrow \tilde y^2-\tilde t^2=a_1 \Longrightarrow
\left(\frac{y'-q'(t)}{y}\right)^2-\left(y-q(t)\right)^2=a_1,
\end{equation}
 and thus the known first integral
(\ref{fintPXVI}) of (\ref{PXVIeq}) is derived. As far as we know
$X^{(\lambda)}$ is a
novel $\lambda$-symmetry of equation (\ref{PXVIeq}).\\

\subsection{Example 4 in \cite{CatalanoMorando09}}
In \cite{CatalanoMorando09} a $\lambda$-symmetry of equation
\begin{equation}
y''=\left( ty'-ty^2+y^2\right)\exp(-1/y)+2\frac{y'^2}{y}+y'\label{ex4}
\end{equation}
 was determined with $\lambda_{k}= t\exp(-1/y)-1/t$.
This equation has been completely solved by considering its solvable structures
\cite{CatalanoMorando09}.

The divergence of  equation (\ref{ex4})-- that has no point symmetries --
yields
\begin{equation} \lambda_J=t\exp(-1/y)+D_t\left(\log\left(y^4\right)+t\right).\end{equation}
If we put $\lambda_{J}$ into the $\lambda$--prolongation
 and solve the determining equations (\ref{Xndeteq}) we get two
$\lambda$-symmetries, i.e.
\begin{equation}
X^{(\lambda)}_1=\frac{1}{y^2\exp(t)}\,\partial_y,\quad\quad
X^{(\lambda)}_2=\frac{1}{y^2\exp(2t)}\left(\frac{1}{y^2}\,\partial_t
+\frac{t}{\exp(1/y)}\,\partial_y\right).
\end{equation}
One can prove that these two $\lambda$-symmetries and that found  in
\cite{CatalanoMorando09} are equivalent. \\The first prolongation of
$X^{(\lambda)}_1$,  i.e.
\begin{equation}
\mbox{pr}X^{(\lambda)}_1=X^{(\lambda)}_1+\left(\frac{2y'}{y^3\exp(t)}+\frac{t}{y^2\exp(t+1/y)}\right)\,\partial_{y'}
\end{equation}
yields the first-order invariants
\begin{equation}  y_1=\frac{y'}{y^2}-t\exp(-1/y),\quad \quad
t_1=t\end{equation} that replaced into equation (\ref{ex4}) generate the
first-order equation
\begin{equation}
y_1 '=y_1 \Longrightarrow y_1=a_1\exp(t) \Longrightarrow
\left(\frac{y'}{y^2}-t\exp(-1/y)\right)\exp(-t)=a_1, \label{fintex4}
\end{equation}
and therefore a first integral of (\ref{ex4}) is derived.

\subsection{Example 5 in \cite{CatalanoMorando09}}
In \cite{CatalanoMorando09} a $\lambda$-symmetry of equation
\begin{equation}
y''=2\frac{y'^2}{y}+\left(t\exp(t/y)-\frac{4}{t}\right)y'
-\left(3\frac{y^2}{t}+y\right)\exp(t/y)+ty^2+2\frac{y}{t^2}\label{ex5}
\end{equation}
 was determined with $\lambda_{k}= t\exp(t/y)$.
  The divergence of  equation
(\ref{ex5})-- that has no point symmetries --  yields
\begin{equation} \lambda_J=t\exp(t/y)+D_t\left(\log\left(\frac{y^4}{t^4})\right)\right).\end{equation}
 If
we put $\lambda_{J}$ into (\ref{Xpr1}) then (\ref{Xndeteq}) yields two
$\lambda$-symmetries, i.e.
\begin{equation}
X^{(\lambda)}_1=\frac{t^3}{y^2}\,\partial_y,\quad\quad
X^{(\lambda)}_2=\frac{t^6}{y^4}\,\partial_t
+\frac{-5t^6y\exp(t/y)+t^8y+5t^5}{5y^3}\,\partial_y.
\end{equation}
One can prove that these two $\lambda$-symmetries and that found  in
\cite{CatalanoMorando09} are equivalent. \\
 The first prolongation of
$X^{(\lambda)}_1$,  i.e.
\begin{equation}
\mbox{pr}X^{(\lambda)}_1=X^{(\lambda)}_1+\frac{t^2}{y^3}\left(t^2y\exp(t/y)-y+2ty'\right)\,\partial_{y'}
\end{equation}
 yields the first-order invariants
\begin{equation}  y_1=\frac{ty^2\exp(t/y)-y+ty'}{ty^2},\quad \quad
t_1=t\end{equation} that replaced into equation (\ref{ex5}) generate the
first-order equation
\begin{equation}
y_1 '=\frac{-3y_1+t^2}{t} \Longrightarrow
y_1=\frac{5a_1+t^5}{5t^3}\Longrightarrow
t^2\frac{ty^2\exp(t/y)-y+ty'}{y^2}-\frac{t^5}{5}=a_1, \label{fintex5}
\end{equation}
and therefore a first integral of (\ref{ex5}) is derived.

\section{Conclusions}

In this paper we have shown that the Jacobi last multiplier provide and
algorithmic simple way to construct $\lambda$--symmetries. Once a
$\lambda$--symmetry is obtained then it is a simple task to derive a first
integral as we have shown in the many examples presented in Section 3.
\\\\
We remark that Strategy 3 as described in \cite{jlm05}
 is not equivalent to the reduction by using $\lambda$-symmetries.
 In fact different output are obtained. For example, equation (\ref{kamkeq})
 was completely solved by quadrature using Strategy 3,
  while the reduction by using
 $\lambda$-symmetries yields just a first integral. Yet those two methods may
 complement each other as in the example of equation (\ref{PXIVpeq}).
\\\\
Also one should be aware of the fact that a $\lambda$-symmetry could be
equivalent to a Lie point symmetry. Lie point symmetries may be obviously
considered $\lambda$-symmetries with $\lambda=0$. For example, let us consider
equation (38) in \cite{MurRo09}, i.e.
\begin{equation} 2yy'' -6y'^2+y^5+y^2=0, \label{38}
\end{equation}
which admits a trivial Lie point symmetry, i.e. $\Gamma=\partial_t$. The
divergence of this equation yields
\begin{equation}
 \lambda_J=6\frac{y'}{y}=D_t(\log(y^6)).\end{equation}
 If we put $\lambda_{J}$ into (\ref{Xpr1}) then (\ref{Xndeteq}) yields one
$\lambda$-symmetry, i.e.
\begin{equation}
X^{(\lambda)}=\frac{1}{y^6}\,\partial_t.
\end{equation}
It is easy to prove that this $\lambda$-symmetry is equivalent to the Lie point
symmetry $\Gamma$.\\
We now search for $\lambda$-symmetries of equation (\ref{VLr2}) which, if
$A=a$, admits two Lie point symmetries, i.e. $\Gamma_1=\partial_t,\, \Gamma_2=
\exp(-at)(\partial_t+a\partial_{r_2})$. The divergence of equation (\ref{VLr2})
yields
\begin{equation}
 \lambda_J=b\exp(r_2) + a. \label{ljVL}\end{equation}
 If we put $\lambda_{J}$ into (\ref{Xpr1}) then (\ref{Xndeteq}) yields one
$\lambda$-symmetry, i.e.
\begin{equation}
X^{(\lambda)}=\exp(-at)\,\partial_{r_2}.
\end{equation}
It is easy to prove that this $\lambda$-symmetry is equivalent to the
Lie point symmetry $\Gamma_2$.\\\\
 A more detailed analysis of the Jacobi last multiplier approach to
 $\lambda$-symmetries is needed. In particular the analysis of higher order
 ODEs can provide new ideas and confirm the importance of this method for the
 integration of ODEs. Moreover its extension to PDEs can provide new insights
 on the meaning of $\mu$-symmetries \cite{CGM04}, \cite{GM04}.
\\\\
 Work is in progress in these directions.

\section*{Acknowledgements}
This work was initiated while DL was enjoying the hospitality of MCN. DL thanks
the Dipartimento di Matematica e Informatica, Universit\`a di Perugia,
 for the provision of facilities.
 DL  has been partly supported by the Italian Ministry of Education and Research,
 2010 PRIN ``Continuous and discrete nonlinear integrable evolutions:
 from water waves to symplectic maps".

\end{document}